# The three-nucleon system as a laboratory for nuclear physics: the need for 3N forces

N. Kalantar-Nayestanaki [1,†] and E. Epelbaum [2,3,‡]

[1] *Kernfysisch Versneller Instituut (KVI), University of Groningen, Zernikelaan 25, 9747 AA Groningen, The Netherlands*
[2] *Forschungszentrum Jülich, IKP (Theorie), D-52425 Jülich, Germany*
[3] *Helmholtz-Institut für Strahlen- und Kernphysik (Theorie), Universität Bonn, Nußallee 14-16, D-53115 Bonn, Germany*
[†] *Electronic address:* nasser@kvi.nl
[‡] *Electronic address:* e.epelbaum@fz-juelich.de

Recent experimental results in three-body systems have unambiguously shown that calculations based on nucleon-nucleon forces fail to accurately describe many experimental observables and one needs to include effects which are beyond the realm of the two-body potentials. This conclusion owes its significance to the fact that experiments and calculations can both be performed with a high accuracy. In this short review, a sample of recent experimental results along with the results of the state-of-the-art calculations will be presented and discussed.

## 1. INTRODUCTION

The ultimate goal of nuclear physics is to understand the properties of nuclei and reactions involving them. Given the smallness of the typical energy scales in nuclear physics, such as the nuclear binding energies (BEs), it appears appropriate to formulate the nuclear N-body problem in terms of the non-relativistic Schrödinger equation. In a first approximation, the two-nucleon potential is sufficient to describe the bulk of the few-nucleon observables at low and intermediate energies. At present, a number of semi-phenomenological two-nucleon models are available which provide an accurate description of the very large nucleon-nucleon (NN) scattering data set below the pion production threshold with a $\chi^2$ per degree of freedom of the order $\approx 1$. Recent advances in the development of few-body methods coupled with a significant increase in computational resources allow, nowadays, to perform accurate microscopic calculations of three- and even four-nucleon scattering observables and of the spectra of light nuclei. This opens the door for precise tests of the underlying dynamics and, in particular, of the role and structure of the three-nucleon force (3NF). One of the simplest and most extensively studied three-nucleon observables is the BE of the triton. It is well known to be significantly underestimated by the existing two-nucleon potentials [1][1]. A similar underbinding occurs for other light nuclei as well [2]. Three-nucleon continuum observables have also been explored by several groups. While the differential cross section of elastic nucleon-deuteron (*Nd*) scattering at low incident beam energies is rather well described using solely two-nucleon potentials, a large discrepancy with the data, known as the $A_y$-puzzle, is observed for the analyzing power. Tensor-analyzing powers and spin-transfer coefficients are generally rather well described at low energy using solely two-nucleon forces (2NFs) but the results of these calculations start to deviate from the data as the energy increases. In addition to the elastic channel, the break-up reaction offers a rich kinematics and as such provides a good testing ground for the structure of the nuclear force.

The observed discrepancies between data and calculations based solely on 2NFs are usually viewed as indication of the existence of a 3NF. Indeed, 3NFs which cannot be reduced to pair-wise NN interactions arise naturally in the context of meson-exchange theory and at the more fundamental level of QCD. At present, several phenomenological 3NF models exist which are typically based on the two-pion-exchange contribution and will be discussed in section II. Despite some remarkable successes of the phenomenological approach, many problems still remain open; see section IV for explicit examples. In addition, there are obvious conceptual deficiencies such as the lack of consistency between the 2NFs and 3NFs. On the other hand, a significant progress in understanding the properties of few-nucleon systems has been achieved recently within the framework of chiral effective field theory (EFT). This approach is linked to QCD via its symmetries and allows to analyze the low-energy properties of hadronic systems in a systematic and controlled way. In addition, it offers a natural explanation for the generally assumed hierarchy of nuclear forces: $V_{2N} \gg V_{3N} \gg V_{4N}$.

In this short review, some of the experimental observables in proton-deuteron (*pd*) scattering will be discussed along with the theoretical developments which are taking place. The experimental investigations have been performed at various laboratories for a large part of the phase space. Here, we restrict ourselves mainly to elastic and break-up observables in the medium energy region of 65 to 200 MeV incident nucleon energy and discuss only a selected set of data. We will also show one example at low energy where there is, as yet, an unresolved problem.

---
[1] Note, however, that phase-equivalent (at low energy) nonlocal NN potentials can be constructed which reproduce the BE of $^3$H.

## II. THEORETICAL FORMALISM

As already pointed out in the introduction, the conventional approach to few-nucleon systems is based on modern semi-phenomenological NN potential models. Typical representatives include the CD-Bonn 2000, Argonne V18 (AV18) and Nijmegen I and II potentials, see [3] for a recent review. In all these models, the (strong) long-range interaction is due to one-pion exchange while the shorter-range contributions are parameterized phenomenologically in different ways. The accuracy of the fit to the NN data-base is comparable with the one of phenomenological partial-wave analyses.

The theory of 3NFs has a long history which goes back to the 30's of the last century, see for example Ref. [4] for an early discussion on the role of the 3NF in nuclear systems. The longest-range contribution due to the two-pion-exchange mechanism is taken into account in all current 3NF models. While the old Fujita-Miyazawa 3NF [5] is governed by the P-wave $\pi N$ scattering amplitude, a more general ansatz based on current algebra is used in the Tuscon-Melbourne 3NF and its chiral-symmetry-corrected version (TM99) [6] as well as in the Brazilian 3NF model [7]. The shorter-range contributions due to the $\pi\rho$ and $\rho\rho$ exchanges were also considered within these models. The Urbana-IX 3NF [8] contains the two-pion-exchange contribution due to the intermediate $\Delta$-excitation and, in addition, a phenomenological shorter-range part. Finally, recent Illinois 3NF models [9] include parameterizations of three-pion-exchange terms due to ring diagrams with one $\Delta$ in the intermediate states.

The usual strategy in few-body studies based on the conventional approach is to adjust the parameter(s) of the 3NF in such a way that the BE of $^3$H is reproduced exactly for a given combination of the 2NF and 3NF models. The resulting nuclear Hamiltonian can be used to compute various few-nucleon observables allowing one to study 3NF effects, see section IV. It should, however, be understood that nuclear potentials are not measurable experimentally and can be modified via an appropriately chosen unitary transformation [10]. 3NF effects should, therefore, always be considered in the context of a particular 2NF model and/or formulation of the problem. For example, in the framework of Ref. [11] based on the extended Hilbert space which includes one $\Delta$, a large portion of what would be the 3NF in a theory without explicit inclusion of $\Delta$ degree of freedom is generated dynamically. The major deficiencies of the conventional approaches are the obvious inconsistency between the few-nucleon forces (2NF, 3NF, ….) and between forces and currents, the lack of systematics in the organization of the dominant dynamical contributions and the poor relation to QCD. It should further be emphasized that the development of the 3NF along the same phenomenological line as in the 2N system, i.e. by parameterizing the most general structure of the 3NF, is not possible (at least at present) due to the much larger number of different possible operators and a much smaller data-base available.

The above difficulties can be overcome in chiral EFT. In this framework, one starts with the most general effective Lagrangian $L_{eff}$ for pions and nucleons consistent with the symmetries of QCD. In particular, pions are identified with the Nambu-Goldstone bosons corresponding to the spontaneously broken $SU(2)_L \times SU(2)_R$ flavour symmetry (i.e. chiral symmetry) of QCD. The approximate chiral symmetry provides strong constraints on the pion interactions. Nuclear forces can be derived from $L_{eff}$ perturbatively via a simultaneous expansion in powers of the low external momenta and about the chiral limit, the so-called chiral expansion [12]. Various techniques have been applied in the past to derive nuclear potentials including e.g. time-ordered perturbation theory, S-matrix based approaches and the method of unitary transformation, see [13] for recent review articles. The importance of a particular contribution to the nuclear force is determined by the power $\nu$ of the expansion parameter $Q/\Lambda$ where Q and $\Lambda$ refer to the generic low-momentum scale associated with external nucleon momenta or $M_\pi$ and the pertinent hard scale, respectively [12]:

$$V = \sum_{(\nu)} V^{(\nu)} = V^{(0)} + V^{(2)} + V^{(3)} + V^{(4)} + \ldots \quad (1)$$

The structure of the nuclear force at lowest orders of the chiral expansion is visualized in Fig. 1. Chiral power counting naturally explains the dominance of the 2NF which starts to contribute at order $\nu$ = 0. At this leading order (LO), the 2NF is given by one-pion-exchange and two NN contact interactions without derivatives. The coupling constants accompanying NN contact interacttions at this and higher orders are determined from the low-energy nucleon-nucleon data. No contribution arises at order $\nu$ = 1 due to parity conservation. The first corrections at next-to-leading order (NLO) with $\nu$ = 2 are due to two-pion exchange, which does not involve new parameters, and contact interactions quadratic in the momenta and $M_\pi$. At next-to-next-to-leading order (N$^2$LO) with $\nu$ = 3 one has to take into account further two-poin-exchange contributions with one insertion of the subleading $\pi\pi NN$ vertices. The corresponding low-energy constants $c_{1,3,4}$ are known from $\pi$N-scattering. In addition, one has first contributions to the 3NF from diagrams as is shown in Fig. 1 [14]. While the two-pion-exchange contribution does not involve new parameters, the two other topologies depend on one unknown constant each, which can be determined from three- or more-nucleon observables. Finally, at next-to-next-to-next-to-leading order (N$^3$LO), $\nu = 4$, one has to account for further corrections to the 2NF which include the sub-subleading two-pion-exchange, the leading three-pion-exchange contributions and the corresponding contact terms. The corrections to the 3NF at N$^3$LO are currently being worked out. In addition, at this order, one has to deal with four-nucleon force (4NF) for the first time. The parameter-free expressions for the leading 4NF at N$^3$LO have been derived recently [15]. As seen in the figure, the hierarchy of $V_{2N} \gg V_{3N} \gg V_{4N}$ is very natural within the framework of EFT. At present, the 2NF has been worked out and applied in the NN system up to N$^3$LO while the results for three and more nucleons are available up to N$^2$LO. For a detailed discussion on the structure and applications of the chiral nuclear forces, the reader is referred to recent review articles [13] and references therein. We further emphasize that alternative schemes for counting the short-range contributions to the nuclear force are currently being explored.

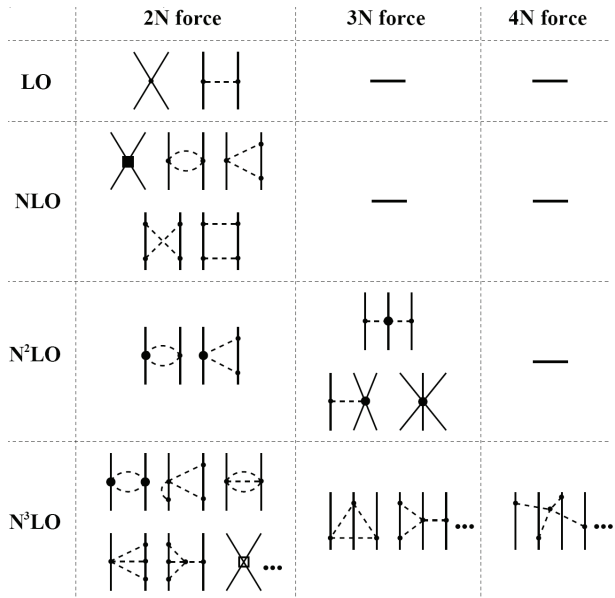

FIG. 1: *Hierarchy of the nuclear forces. Solid (dashed) lines represent nucleons (pions). Solid dots, filled/open squares and filled circles depict vertices from $L_{eff}$.*

Theoretical predictions for 3N scattering observables presented in this work have been obtained by solving the corresponding Faddeev equations, see [16] for the details. The inclusion of the long-range electromagnetic interaction requires a non-trivial generalization of the formalism, see [16] and references therein and [17] for recent important numerical progress on that front.

## III. EXPERIMENTAL TECHNIQUES

The experimental equipment in the study of the three-body systems depends on the reaction and the observable one is considering. For example, in the study of the cross section of elastic *pd* scattering, one needs an unpolarized source of protons or deuterons and a small detector which is capable of identifying one of the outgoing particles with a high energy/angle resolution. A magnetic spectrometer with a small solid angle would be enough for this purpose. If one is interested in the analyzing power of the reaction, one would need, in addition, a polarized source of particles. In the study of spin-correlation coefficients (such as those performed at IUCF), one would need a polarized target as well, and the investigation of spin-transfer coefficients (such as those at RIKEN and KVI) would require, instead, the measurement of the polarization of the outgoing particles. For the break-up studies, the detection system becomes more complicated since the reaction includes three particles in the final state. In this case, one could either choose a very specific geometry in the measurement or employ a detector with a very large acceptance in order to accommodate the detection of more than one particle in a large part of the reaction phase space. An example of such a device, as shown in Fig. 2, is the BINA detector recently commissioned at KVI. The design of recent detectors such as the one shown in Fig. 2, is based on the tremendous experience gained with the pioneering experiments performed at lower energies at Cologne, Bonn, Durham (TUNL) and Kyushu. In this article we, however, focus mainly on intermediate energies.

All the interactions discussed here are hadronic interactions with generally rather large cross sections. Therefore, detectors should be capable of dealing with high rates. This, in turn, reduces the required amount of beam time in general. However, in some regions of phase space where the cross sections are small, the measurement of spin observables requires long running times. In the past, solid $CH_2$ or $CD_2$ targets were used. With the advent of thin and strong windows, liquid targets have become an alternative, introducing also less background.

## IV. RESULTS AND DISCUSSION

As already pointed out in the introduction, an important indication of the significance of the 3NF effects comes from the BEs of light nuclei. The addition of the 3NF removes the observed underbinding by the 2NFs of typically more than 0.5 MeV (4 MeV) for $^3$H ($^4$He) and the bulk of discrepancy for heavier nuclei [2].

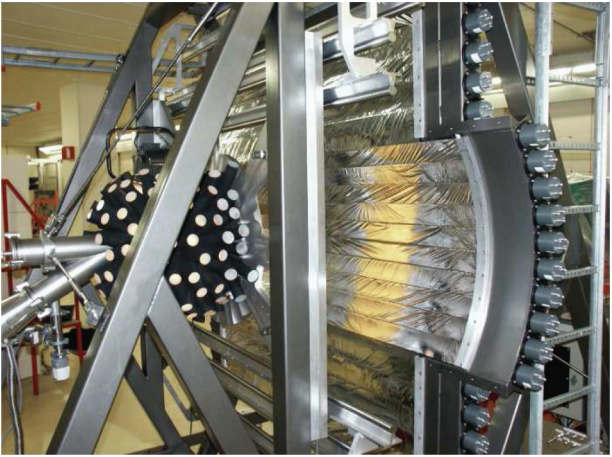

FIG. 2: *A picture of the Big Instrument for Nuclear-polarization Analysis (BINA) at KVI. One can see the forward wall of the detector along with the backward ball which also functions as the vacuum chamber. The wire chamber and the thin ΔE detector which are placed between the ball and the forward wall, and the phototubes of the ball along with all the cables are taken out in this picture.*

On the other hand, there are also low-energy 3N scattering observables, for which the addition of the available 3NFs does not improve the description of the data. Well-known examples are the low-energy (below 10 MeV) $Nd$ analyzing powers ($A_y$-puzzle) [16].

At intermediate energies ($\gtrsim$ 65 MeV/nucleon), calculations have shown that effects due to 3NF should clearly manifest themselves in the minimum of the differential cross section of the elastic channel [18]. While the contribution of the 3NF to the differential cross section is rather constant across the center-of-mass angular range, the contribution from pure two-nucleon interactions has a pronounced dip around 90° c.m. angle and decreases relative to the contribution of the 3NF as the incident-beam energy increases. In the literature, there exist some measurements of cross sections and analyzing powers of the elastic channel at intermediate energies [19-25]. However, most of them either suffer from a limited angular coverage or lack the accuracy needed to compare with modern calculations. Recent measurements have made an effort to improve on both fronts. Aside from KVI data [26,27], results have been obtained at RIKEN, RCNP, IUCF and Saturne at intermediate energies. Here, we only present a part of the experimental results for intermediate energies.

In Fig. 3, the results of cross sections and analyzing powers are shown by open squares as a function of the center-of-mass angle for a number of measurements performed at KVI at energies ranging from 100 to 200 MeV [26,27]. Data available in the literature at nearby energies are also shown. In all cases there seems to be a good agreement between various data sets. The only exception is the cross sections at 135 MeV, where the KVI and the RIKEN data [28] do not agree with each other. Results of calculations based on different modern 2N potentials are shown by the black band, whose width can be regarded as a measure of model dependence. In addition, results are shown for various 2NFs combined with TM99 3NF (grey band) and AV18 potential combined with Urbana-IX 3NF (solid line). These results also agree with coupled-channel calculations by the Hanover group [11]. It is clearly seen that the calculations based on 2NFs alone are capable of describing the bulk of the cross sections for two orders of magnitude variation in this observable. However, major discrepancies set in especially in the cross section minima. The discrepancies also increase as a function of incident beam energy. The addition of various 3NFs brings the results closer to the data but does not remove the discrepancies at backward angles, which also grow with increasing incident beam energy. The analyzing powers show a similar behavior, namely that the disagreements with the data decrease when adding the three-body force but remain at the backward angles. Note that relativistic corrections have been found to be negligible for these observables at the energies considered [29]. It should also be emphasized that the experiments have been carried out with (polarized) proton beams while the calculations presented here neglect the effect of the Coulomb force. The Coulomb effect has indeed been shown recently to be very small in this particular case [17]. The calculations based on chiral EFT have so far been carried out for $Nd$ scattering up to $N^2LO$ in the chiral expansion. For these relatively high energies, the theoretical uncertainty at $N^2LO$ appears to be rather large, making the comparison with data not particularly useful. It is important to extend this analysis to $N^3LO$, where one expects smaller theoretical uncertainty. This work is in progress.

The results of the cross sections and analyzing powers in the elastic channel indicate that the deficiencies in the existing three-body forces show up not only in their central but also in the spin part. In order to investigate these aspects, more experiments have been performed at KVI and RIKEN. These difficult measurements involved not only incoming beam polarizations, but also the polarization of the outgoing particles has been measured in the elastic channel. The observables are called spin-transfer coefficients. In Fig. 4, results of recent measurements at KVI [30] are shown. These measurements were done with a 180 MeV polarized deuteron beam. The measurement of the polarization of the outgoing particle results in several observables all shown in the figure. In addition to cross sections and vector and tensor analyzing powers ($A_y$ and $A_{yy}$ which only involve a polarized beam), we now have induced polarization ($P_y$, which is produced by an unpolarized beam) and vector and tensor spin-transfer coefficients ($K_y^{y'}$ and $K_{yy}^{y'}$, which contain information about the polarization transfer from the incident deuteron beam to the ejectile proton). For the sake of simplicity, calculations based solely on 2NFs are not shown. At this energy, the effect of 3NFs is smaller than for the cases shown in Fig. 3. In Fig. 4, the results for the AV18 potential combined with Urbana-IX 3NF (dashed curve) and the coupled-channel calculation by the Hanover group including the Δ (solid curve) are shown together with the results of the chiral EFT at N$^2$LO (gray band). The band serves as an estimation of the theoretical uncertainty at this order. The results for the cross section in the chiral EFT approach show some deviations from the data. However, given the size of the theoretical uncertainty at this order, the chiral results are in good agreement with the data. Clearly, one needs to extend the calculations to N$^3$LO in order to reduce the uncertainty and to be able to analyze the data shown in Fig. 3.

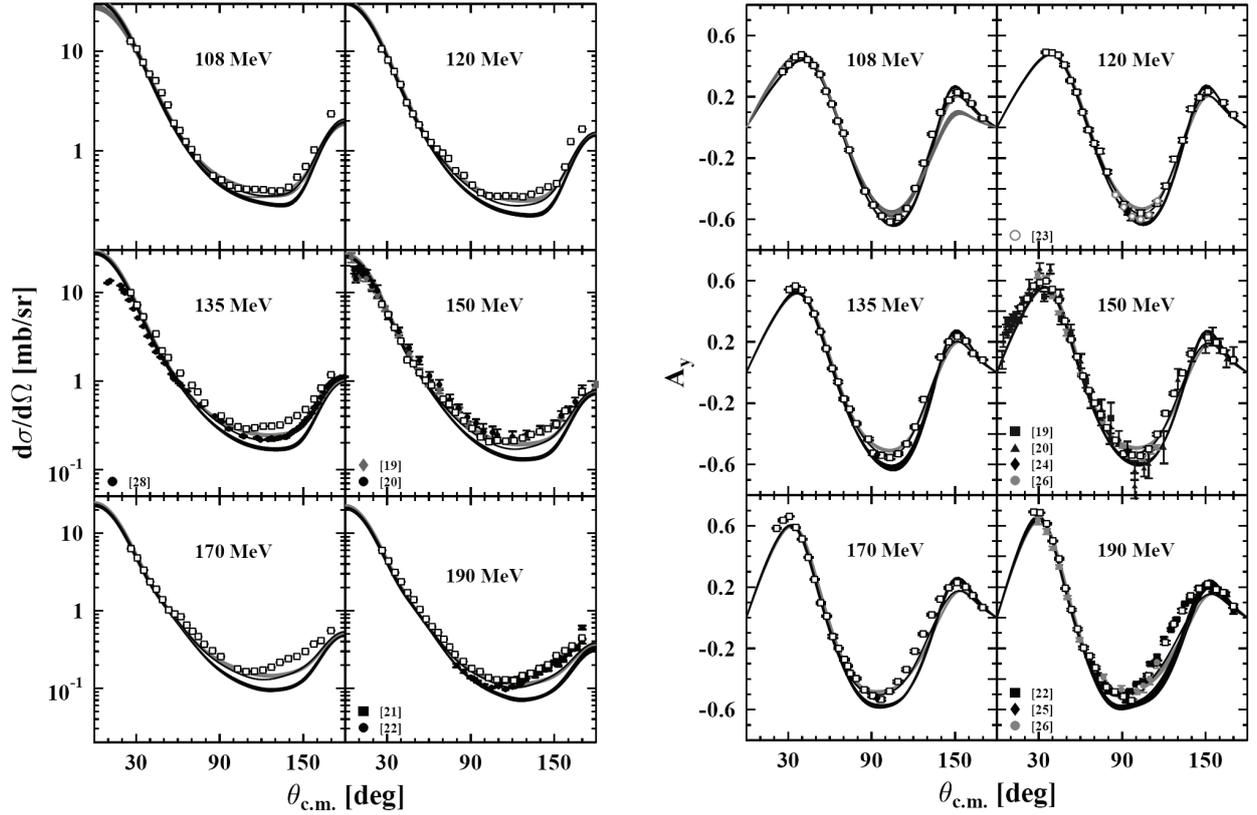

FIG. 3: *Differential cross sections (left) and vector analyzing powers (right) for elastic pd scattering versus the scattering angle for different incident-beam energies at KVI (open squares). Also shown are the available data at nearby energies and calculations from NN potentials (black band), NN+TM99 (gray band), AV18+Urbana-IX (solid line). The solid line overlaps in most places with the gray band. For each data point, only the statistical uncertainty is given.*

The *Nd* break-up reaction offers even more possibilities than the elastic channel due to the much richer kinematics corresponding to three nucleons in the final state. It requires, however, more complicated experiments. Measurements at low energies have been performed in the past at Bonn, Cologne, TUNL and Kyushu producing interesting and sometimes surprising results. In particular, the so-called symmetric space-star configuration (SST) appears rather puzzling. In this configuration, the plane in the c.m. system spanned by the outgoing nucleons is perpendicular to the beam axis, and the angles between the nucleons are 120°. At $E_{lab} =$ 13 MeV, the proton-deuteron and neutron-deuteron (*nd*) cross section data deviate significantly from each other. Theoretical calculations based on both phenomenological and chiral nuclear forces have been carried out for the *nd* case and are unable to describe the data. In addition, the Coulomb effect was found to be far too small to explain the difference between the *pd* and *nd* data. Recently, proton-deuteron data for a similar symmetric constant relative-energy (SCRE) configuration have been measured in Cologne [31] at $E_d = 19$ MeV. This geometry is characterized by the angle α between the beam axis and the plane in the c.m. system spanned by the outgoing nucleons. Similar to the SST geometry, one observes large deviations between the theory and the data, in particular for α = 56°, see Fig. 5. The included 3NFs have little effect on the cross section while the effect of the Coulomb interaction is significant but removes only a part of the discrepancy.

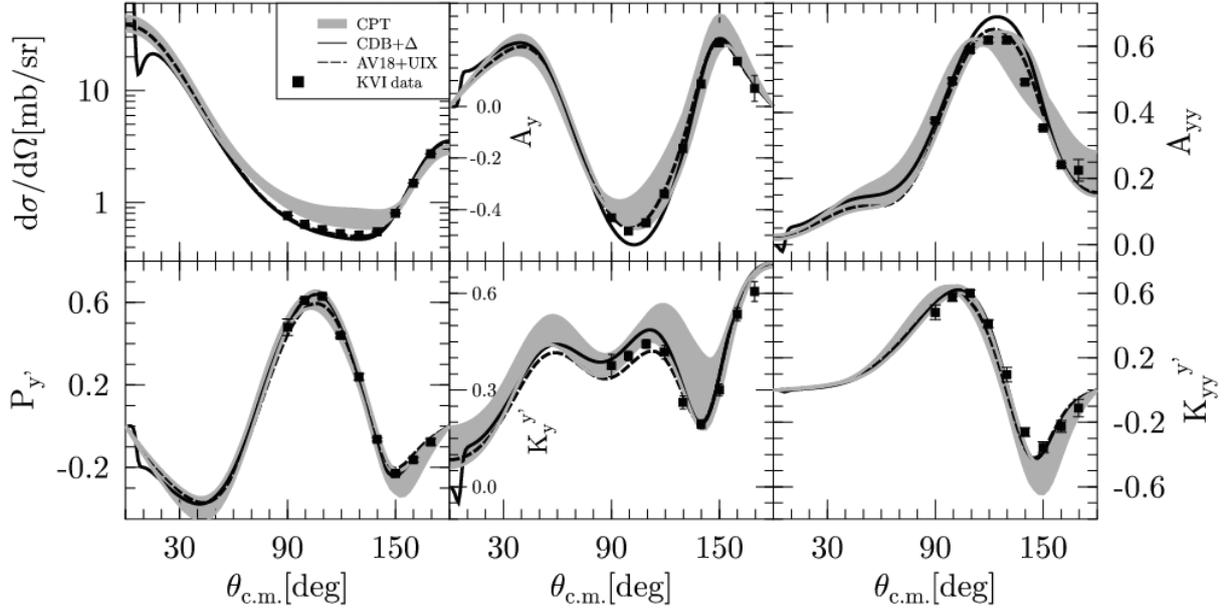

FIG. 4: *Results for the cross sections, vector and tensor analyzing powers, and vector and tensor spin-transfer coefficients of the deuteron-proton elastic scattering at 90 MeV/nucleon beam energy in comparison with various models. The gray band shows the results of the chiral EFT calculations at $N^2LO$, the solid and dashed curves show the results based on the coupled-channel calculations including the explicit $\Delta$ and from the AV18+Urbana-IX potential, respectively. Only statistical uncertainties are shown here.*

At intermediate energies, first attempts have been made to perform break-up experiments in which a large part of the phase space is covered simultaneously. This requires detection setups with a large solid angle coverage. The first results of such measurements performed at KVI at 65 MeV/nucleon have been published [32]. Even though the effect of 3NFs are not large at this energy, a $\chi^2$ analysis of more than 1500 data points reveals that the inclusion of the three-body forces certainly improves the agreement between the theoretical predictions and the data. Chiral EFT results at $N^2LO$ are of a comparable quality to the ones based on modern phenomenological nuclear forces. As was mentioned above, the Coulomb force can now be included exactly in the *pd* break-up calculations. With the large variety of kinematical variables in the break-up reaction, one can choose such kinematical conditions that the relative energy of the two outgoing protons is small, which enhances the Coulomb effect. Figure 6 shows a small fraction of the data measured at KVI. Here, one can clearly see a nice agreement between the theory and data and the radical improvement in some configurations after the inclusion of the Coulomb force. These benchmark studies demonstrate that various ingredients needed to study 3NF effects are now under control while the currently available 3NF models still suffer from severe deficiencies. Clearly, more theoretical input is needed in order to understand the details of the structure of the 3NF and to find out where the deficiencies actually lie.

### V. SUMMARY AND CONCLUSIONS

In this article, a sample of recent high-precision data at intermediate energies for various observables and different reaction channels and recent data at a low energy for a particular geometry in the break-up reaction were presented along with various theoretical calculations. Over a large energy and angle range, it was shown that the bulk of the data is described by modern calculations especially after the inclusion of the 3NF. Since both the data and the calculations have relatively small uncertainties, definite conclusions can be drawn from the comparisons. In particular, the cross-section data for the elastic channel seem to be well described at lower energies by calculations including 3NFs. Discrepancies set in as one increases the incident beam energy. This could be an indication that relativistic effects are setting in and/or three-nucleon forces used in the calculations are still not quite under control. First studies indicate, however, that in the elastic channel, relativistic effects are very small to the highest energy shown here.

For the proton analyzing powers in the elastic channel, calculations based on different 3NFs not only deviate from the data but even differ from each other.

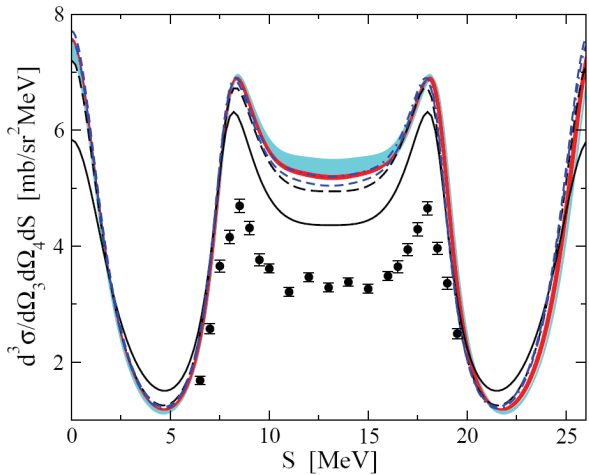

FIG. 5: *Results for the SCRE deuteron break-up cross sections at $E_d$= 19 MeV as a function of the kinematical variable S, with $\alpha$ = 56°. Light (dark) shaded bands depict chiral EFT predictions at NLO ($N^2LO$). Short-dashed and solid (long-dashed) lines show the results based on the combination of the CD-Bonn 2000 2NF and TM99 3NF and the coupled-channel calculations including the explicit $\Delta$ with (without) Coulomb interaction. The results based on the CD-Bonn 2000 2NF overlap with the chiral EFT band at $N^2LO$.*

The deviations seem to increase as a function of energy. The spin structure of 3NF needs further attention. This can be seen also from the analyzing powers and spin-transfer coefficients, using a deuteron beam at lower energy, which bring forward a mixed message depending on the observable. This trend has also been observed in recent measurements performed at RIKEN, Japan.

In general, it can be concluded that three-nucleon observables, like $d\sigma/d\Omega$ or $A_y$, are well described by modern NN potentials combined with phenomenological 3NFs at low intermediate energies (above 30 MeV). However, for higher energies and for backward angles, deviations from data are observed. This calls for further theoretical investigations.

For the break-up reaction, new results have emerged showing the need for the inclusion of 3NFs. The Coulomb force has recently been included in the break-up calculations and leads to substantial effects in some configurations. More experiments in this channel have been done for a range of incident beam energy and results are emerging.

The calculations within chiral EFT have been performed for the lowest energies considered in this work (up to 90 MeV) and up to $N^2LO$. This framework is systematically being improved and allows for a consistent treatment of the 2NFs and 3NFs. It is important to extend the calculations to $N^3LO$ in order to test the convergence of the chiral expansion and to increase the predictive power for a large energy range. The parameter-free expressions for the 3NF at this order are currently being worked out.

## Acknowledgements

The Bochum-Cracow and Hanover-Lisbon groups are acknowledged for providing the results of their calculations for this article. We also like to acknowledge K. Ermisch, H.R. Amir-Ahmadi, H. Mardanpour, J.G. Messchendorp, St. Kistryn and the Polish colleagues from Cracow and Katowice for the work on the experiments, parts of which are presented here, as well as W. Glöckle, M.N. Harakeh, U.-G. Meißner, A. Nogga and R.G.E. Timmermans for useful comments on the manuscript.

## REFERENCES

[1] A. Nogga et al., Phys. Rev. Lett. 85 (2000) 944.
[2] S.C. Pieper and R.B. Wiringa, Ann. Rev. Nucl. Part. Sci. 51 (2001) 53.
[3] R. Machleidt, I. Slaus, J. Phys. G27 (2001) R69.
[4] H. Primakoff and T. Holstein, Phys. Rev. 55 (1939) 1218.
[5] J. Fujita and H. Miyazawa, Prog. Theor. Phys. 17 (1957) 360.
[6] S.A. Coon and H.K. Han, Few Body Syst. 30 (2001) 131.
[7] H.T. Coelho et al., Phys. Rev. C28 (1983) 1812.
[8] B.S. Pudliner et al., Phys. Rev. C56 (1997) 1720.
[9] S.C. Pieper et al., Phys. Rev. C64 (2001) 014001.
[10] W. Glöckle and W. Polyzou, Few-Body Syst. 9 (1990) 97.
[11] A. Deltuva et al., Phys. Rev. C68 (2003) 024005.
[12] S. Weinberg, Phys. Lett. B251 (1990) 288; Nucl. Phys. B363 (1991) 3.
[13] P.F. Bedaque and U. van Kolck, Ann. Rev. Nucl. Part. Sci. 52 (2002) 339; E. Epelbaum, Prog. Part. Nucl. Phys. 57 (2006) 654.
[14] U. van Kolck, Phys. Rev. C49 (1994) 2932; E. Epelbaum et al., Phys. Rev. C66 (2002) 064001.
[15] E. Epelbaum, Phys. Lett. B639 (2006) 456.
[16] W. Glöckle et al., Phys. Rept. 274 (1996) 107.
[17] A. Deltuva et al., Phys. Rev. C71 (2005) 064003.
[18] H. Witała et al., Phys. Rev. Lett. 81 (1998) 1183; S. Nemoto et al., Phys. Rev. C58, (1998), 2599.
[19] H. Postma and R. Wilson, Phys. Rev. 121 (1961) 1229.
[20] K. Kuroda et al., Nucl. Phys. 88 (1966) 33.
[21] G. Igo et al., Nucl. Phys. A195 (1972) 33.
[22] R.E. Adelberger and C.N. Brown, Phys. Rev. D5 (1972) 2139.
[23] S.P. Wells et al., Nucl. Inst. Meth. Phys. Res. A 325 (1993) 205.
[24] E.J. Stephenson et al., Phys. Rev. C68 (2003) 051001(R).
[25] R.V. Cadman et al., Phys. Rev. Lett. 86 (2001) 967.
[26] R. Bieber et al., Phys. Rev. Lett. 84 (2000) 606.
[27] K. Ermisch et al., Phys. Rev. C71 (2005) 064004.
[28] K. Sekiguchi et al., Phys. Rev. Lett. 95 (2005) 162301.

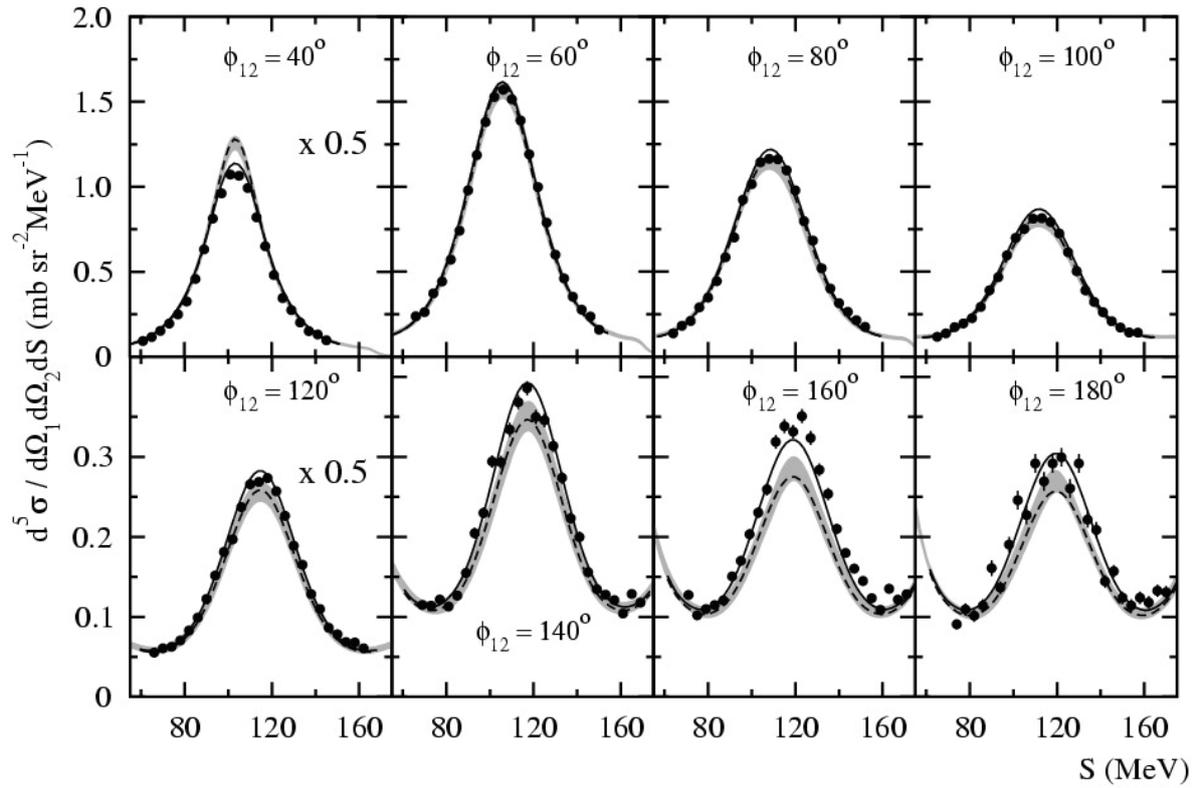

FIG. 6: *Results for the break-up cross sections for various coplanarity angles between the two outgoing protons ($\varphi_1 - \varphi_2$) as a function of the kinematical variable S, for an incident deuteron energy of 130 MeV and the outgoing proton polar angles of (15°, 15°). The band shows the results of the chiral EFT calculations at $N^2LO$. The curves are the results of the coupled-channel calculations including the explicit $\Delta$ (dashed curve) and also including the Coulomb force (solid curve).*

[29] H. Witała, Phys. Rev C71 (2005) 054001.
[30] H.R. Amir-Ahmadi et al., Phys. Rev. C, in press.
[31] J. Ley et al., Phys. Rev. C73 (2006) 064001.
[32] St. Kistryn et al., Phys. Rev. C72 (2005) 044006; Phys. Lett. B641 (2006) 23.